\documentclass[twocolumn,prd,superscriptaddress,aps,nofootinbib,floats,floatfix,amsmath,amssymb,secnumarabic]{revtex4}
\usepackage[dvips,final]{graphicx}
\usepackage{amsmath}  
\usepackage{graphicx} 
\usepackage{fancybox} 
\usepackage{mathrsfs} 
\usepackage{amssymb} 
\usepackage{setspace}
\usepackage{verbatim}
\usepackage{subfigure}

\newcommand{\pdiff}[2]{\frac{\partial{#1}}{\partial{#2}}}

\newcommand{\diff}[2]{\frac{d{#1}}{d{#2}}}

\newcommand{\vl}{\emph{Via Lactea II} }

\newcommand{\be}{\begin{equation}}
\newcommand{\ee}{\end{equation}}
\newcommand{\ba}{\begin{array}}
\newcommand{\ea}{\end{array}}
\newcommand{\bqa}{\begin{eqnarray}}
\newcommand{\eqa}{\end{eqnarray}}

\renewcommand{\d}{\mathrm{d}}

\newcommand{\bsh}{B_{\rm\scriptscriptstyle SH}}
\newcommand{\bmh}{B_{\rm\scriptscriptstyle MH}}
\newcommand{\mchi}{M_{\rm\scriptscriptstyle DM}}

\begin{document}
$\phantom{.}$
\vskip-1.5cm
\rightline{CERN-PH-TH/2010-011}

\title{Leptons from Dark Matter Annihilation in Milky Way Subhalos}

\author{James M. Cline\footnote{jcline@hep.physics.mcgill.ca}}
\affiliation{Theory Division, CERN, CH-1211, Gen\`eve, Switzerland} 
\affiliation{Department of Physics, McGill University,
3600 Rue University, Montr\'eal, Qu\'ebec, Canada H3A 2T8}

\author{Aaron C. Vincent\footnote{vincenta@hep.physics.mcgill.ca}}
\affiliation{Department of Physics, McGill University,
3600 Rue University, Montr\'eal, Qu\'ebec, Canada H3A 2T8}

\author{Wei Xue\footnote{xuewei@hep.physics.mcgill.ca}}
\affiliation{Department of Physics, McGill University,
3600 Rue University, Montr\'eal, Qu\'ebec, Canada H3A 2T8}


%
%

\date{\today}


\begin{abstract}

Numerical simulations of dark matter collapse and structure formation
show that in addition to a large halo surrounding the baryonic
component of our galaxy, there also exists a significant number of
subhalos that extend hundreds of kiloparsecs beyond the edge of the
observable Milky Way. We find that for dark matter (DM) annihilation
models, galactic subhalos can significantly modify the spectrum of
electrons and positrons as measured at our galactic position. Using
data from the recent \vl simulation we include the subhalo
contribution of electrons and positrons as boundary source terms for
simulations of high energy cosmic ray propagation with a modified
version of the publicly available GALPROP code. Focusing on the DM DM
$\rightarrow$ 4e annihilation channel, we show that  including
subhalos leads to a better fit to both the Fermi and PAMELA data. 
The best fit gives a dark matter particle mass of 1.2 TeV, for boost
factors of  $\bmh = 90$ in the main halo and $\bsh = 1950-3800$
in the subhalos (depending on assumptions about the background),  in contrast to the  0.85 TeV mass that gives the best fit in
the main halo-only scenario.   These fits suggest that at least a
third of the observed electron cosmic rays from DM annihilation could
come from subhalos, opening up the possibility of a relaxation of
recent stringent constraints from inverse Compton gamma rays
originating from the high-energy leptons. 
\end{abstract}

\maketitle


\section{Introduction and Summary}

Recent observations of the spectrum of electrons and positrons by the
Fermi collaboration \cite{fermi} and of the positron fraction
$e^+/(e^++e^-)$ by the PAMELA experiment \cite{pamela} hint at a
possible new source of cosmic ray $e^+$ and $e^-$ in the TeV energy
region. According to recent models
\cite{ArkaniHamed:2008qn,Cholis:2008qq,Cholis:2008wq,Meade,Chen:2009ab,
Watson, Kuhlen:2008aw}, these excesses could be the signal of dark
matter (DM) annihilation via a dark sector gauge boson that allows a
Sommerfeld-type enhancement at low velocities. Best fits to the
electron-positron spectra indicate that the dark matter candidate
$\chi$  that annihilates within the galaxy should have a mass of
around $\mchi \simeq$ 1 TeV, and annihilate into two pairs of light
leptons via the process DM DM $\rightarrow \phi\phi \rightarrow 4e$
or DM DM $\rightarrow \phi\phi \rightarrow 4\mu$. The particle $\phi$
should furthermore be light enough not to decay into $p \bar p$ pairs
since excess antiprotons are not observed by PAMELA.\footnote{however
see ref.\ \cite{Watson} for arguments that this constraint may not be
necessary, due to astrophysical uncertainties in the background
model}

While the visible galaxy spans a diameter of approximately 40 kpc and
a height of 8 kpc, $N$-body simulations \cite{vl2, aquarius} predict
a roughly spherical structure of dark matter subhalos whose peak
concentration occurs $\sim$ 70 kpc from the galactic center (GC) and
extends as far as several thousand kpc. Relative velocities between
particles in these regions are one to two orders of magnitude smaller
than in the Milky Way's main halo, and the relative overdensity of
such regions make them ideal sources of DM annihilation products.
This has been explored by other authors in the context of gamma ray
signals originating from subhalos
\cite{Bovy,Kuhlen:2009jv,Kistler:2009xf,Ando:2009fp,
kuhlen,Kuhlen:2008aw} and found to be significant. These gamma rays,
which originate from final-state radiation, are not the main product
of this class of DM annihilation; rather they are by-products of 
charged particles and neutral pions.  

In this work we consider the possibility that the excess leptons
observed by PAMELA and Fermi/LAT themselves have a strong component
originating in the subhalos. This possibility was previously
considered in ref.\ \cite{Brun:2009aj}, but there it was assumed that
one or two nearby subhalos would dominate any additional contribution
to the signal. Here we will show that the best fits to the
data are found by taking into account the full ensemble of
substructures.  It will be seen that the subhalos that individually
contribute weakly to the lepton flux are nevertheless so numerous
that their combined effects cannot be neglected.

We used a modified version of the GALPROP cosmic ray propagation
code, in which leptons from distant subhalos give a new source term
at the boundary of the diffusion zone.  The data of the \vl
simulation \cite{vl2} are taken as our model for the subhalos. We allow for
independently adjustable boost factors for the main halo and
subhalos, motivated by the fact that Sommerfeld enhancement of the
annihilation cross section can be much greater in the  subhalos due
to their lower velocity dispersion \cite{ArkaniHamed:2008qn}.  If we
also allow the background electron and positron flux normalizations
to be rescaled, as in references
\cite{Cirelli:2009dv,Meade,Papucci:2009gd}, we find that the
inclusion of subhalos gives a much better fit to both cosmic ray data
sets, with the best-fit DM particle mass of  $\mchi = 1.2$ TeV.  

On the other hand, if the $e^+$ and $e^-$ backgrounds are  instead
fixed at the GALPROP output level, it is known that there is a
discrepancy between the boost factors needed for explaining PAMELA
and Fermi, even in the standard main halo-only scenario. This
discrepancy remains in the subhalo scenario, where we find that a DM
mass of $\mchi =2.2$ TeV improves the fit to the Fermi data,  whereas
the fit to PAMELA is not improved. 

Our results suggest that the inclusion of leptons from DM
annihilation in the subhalos surrounding the Galaxy should affect not
only the amplitude, but also the shape of the observed spectrum due
to inverse Compton scattering (ICS) of the leptons with the radiation
fields inside the observable galaxy.  This is potentially important
because the most recent constraints on this effect
\cite{Papucci:2009gd, Cirelli:2009dv} effectively rule out the DM
annihilation interpretation of the Fermi excess, and leave only a
very reduced corner of parameter space consistent with PAMELA.  We
hope to quantitatively address the question of whether substructure
indeed allows for relaxation of these constraints in the near future.

In section 2 we briefly describe GALPROP and our choices of parameters
for cosmic ray propagation.  Section 3 details the modifications we
made to GALPROP in order to include the $e^+ e^-$ pairs from DM
annihilation in the subhalos.  The results are presented in section
4, and conclusions in section 5.


\section{Cosmic ray propagation models}
\label{crm}

Inside the diffusive zone of the Galaxy, 
cosmic ray species propagate according to the transport equation
\cite{Strong:1998pw}
\begin{eqnarray}
\label{diffeqn}
\pdiff{\psi}{t} & = & q(r,z,p) + \nabla \cdot (D_{xx} \nabla \psi - \vec V_c \psi )  \\
		& & + \frac{\partial}{\partial p} p^2 D_{pp}
\frac{\partial}{\partial p}\frac{1}{p^2} \psi 
		-\frac{\partial}{\partial p}\left[ \diff{p}{t}
\psi - \frac{p}{3}(\nabla \cdot \vec V_c) \psi \right].\nonumber
\end{eqnarray}
$\psi(\vec{x},p,t)$ is the particle density per unit momentum $p
\equiv|\vec p|$,
$q(\vec{x},p)$ is the source term, $D_{xx}$ is the energy-dependent
diffusion coefficient, $D_{pp}$ quantifies reacceleration via
diffusion in momentum space and $\vec{V_c}$ is the convection
velocity. In the case of composite species, terms accounting for
radiative decay and fragmentation must furthermore be included. For
$D_{xx}$ we use the parametrization \cite{Simet:2009ne}
\begin{equation}
 D_{xx} = D_{0xx}\left(\frac{E}{{4\ \rm GeV}}\right)^\delta,
\end{equation} 
where $E$ is the particle energy and 
$D_{0xx}$ is the diffusion coefficient at 
reference energy $E = 4$ GeV.\footnote{More precisely, diffusion
depends on particle rigidity, the energy divided by the charge.  We
assumed the particles have unit charge here.}\ \ 
$D_{0xx}$ and the exponent $\delta$ are determined by 
fitting to heavy nuclei cosmic ray data.

There are two widely-used approaches to cosmic ray propagation within the
galaxy. The first is a semi-analytic model in which the baryonic
component of the galaxy is accelerated in a thin disk at $z = 0$ from
which particles diffuse according to a Bessel series expansion until
$z = \pm L_{\rm eff}$, beyond which they freely escape. The second,
fully numerical, approach implemented in the publicly available
GALPROP \cite{Strong:1998pw} package uses a Crank-Nicholson scheme to
solve eq.\ (\ref{diffeqn}) within a diffusion zone of height $L_{\rm eff}$ and
radius $R_{\rm eff}$. An advantage of the latter technique is that it
allows the use of realistic maps of radiation and gas in the
propagation scheme. While this is not the focus of our paper, it is
relevant to point out that differences between models are responsible
for differences between fits in recent dark matter annihilation
models. This discrepancy has been known for some time; see for
example the discussion in ref.\ \cite{Maurin:2002hw}. Nevertheless,
we shall henceforth focus exclusively on the numerical approach. Our
simulations were run using a modified version of GALPROP 50.1p that
was graciously provided by the authors of ref.\ \cite{Cholis:2008qq}
and which we further modified to handle subhalo sources.

The strongest available constraints on cosmic ray propagation models
are ratios of secondary-to-primary species such as B/C or sub-Fe/Fe.
The authors of ref.\ \cite{Simet:2009ne} conducted an exhaustive
search of the GALPROP parameter space for input values that gave best
fits to 12 secondary/primary cosmic ray experiments. For our
simulation runs we took their best fit parameters: $D_{0xx}$ = 6.04
$\times$ $10^{28}$ cm$^2$ s$^{-1}$ (0.19 kpc$^2/$Myr), $L_{\rm eff} = 5.0$ kpc, $\delta
= 0.41$, with no convection. We used an Alfv\'en speed $V_A = 31$ km
s$^{-1}$, which gave a slightly better fit to the HEAO B/C data
\cite{Engelmann:1990zz}. It should be noted that these parameters are
quite different from the corresponding best fits of the semi-analytic
model used, for example by Meade et al. \cite{Meade}. 



\section{Including subhalo flux in GALPROP}
\label{subhalo}

The many-body simulation \vl \cite{vl2}, which modeled the
evolution and collapse of more than $10^9$ particles over the history
of a Milky Way-sized structure, resolves over 20,000 dark matter
subhalos around the galactic host halo.   The data characterizing
each of these subhalos is publicly available \cite{vl2data}.
While the visible galaxy is
only some 40 kpc across, these subhalos extend as far out as 4000 kpc
from the galactic center. Each subhalo is locally
much denser than the host halo and has its own  radial velocity
dispersion profile. The annihilation rate of dark matter,
proportional to $\rho^2$, should thus spike within these subsystems
when compared to the annihilation rate of diffuse DM particles of the
host halo. 

For a given subhalo $i$ at a distance $\ell_i$ from the edge of the
diffusion zone of the galaxy, the flux of $e^+$ or $e^-$ 
on this boundary takes the form
\be
	{d\Phi_i\over dE}  = 
	 \bsh \langle \sigma v\rangle\,  {dN\over dE}(\ell_i) \int_0^\infty {r^2\rho_i^2\over
 \ell_i^2 \mchi^2}\, dr\,
\label{flux}
\ee
where $\bsh$ is an average boost factor for the subhalos due to 
Sommerfeld enhancement 
for example, and $\rho_i(r)$ is the mass density profile of the
subhalo.   The unboosted cross section is assumed to be
$\langle \sigma v \rangle =  3 \times 10^{-26}$ cm$^3$ s$^{-1}$ 
in accordance with the standard assumption that the DM abundance
was determined by freeze-out starting from a thermal density.
In a more exact treatment, the boost factor would be
velocity  dependent \cite{MarchRussell:2008tu,Lattanzi:2008qa} and appear within the average over DM velocities
indicated by the brackets in $\langle \sigma v\rangle$.   Moreover
each subhalo in general has a different boost factor since the
velocity dispersions that determine $\bsh$ depend on the size of the
subhalo \cite{Bovy}. For this preliminary study, we simply parametrize the effect
by an average boost factor, where the averaging includes the sum over
all subhalos as well as the integration over velocities.  

The energy spectrum $dN/dE$ of electrons from the DM annihilations is
taken for simplicity to be a step function at the interaction point,
$dN/dE = \mchi^{-1}\Theta(\mchi-E_0)$, where $E_0$ is the energy
immediately following the annihilation.  We are interested in models
where the DM particles initially annihilate into two hidden sector
gauge or Higgs bosons, each of which subsequently decays into
$e^+e^-$ \cite{ArkaniHamed:2008qn}.   The four-body phase space would
thus be a more exact expression for $dN/dE$, but the step function
has the correct qualitative shape and is simpler to implement in
GALPROP.

The energy of the electron at the edge of the galaxy is reduced from
its initial value $E_0$ by scattering with CMB photons before
reaching the galaxy (starlight, infrared radiation and synchrotron
radiation are only important in the inner galaxy
\cite{Porter:2005qx}), according to the loss equation $dE/d\ell =
-\kappa E^2$  \cite{Meade} where $\kappa = ({4 \sigma_T }/{3
m_e^2})\, u_{\mathrm{CMB}}  = 6.31 \times 10^{-7} \mathrm{\ kpc\
GeV}^{-1}$, $\sigma_T = \frac{8\pi}{3}(\alpha_{EM}\hbar/m_e c)^2$ is the 
Thomson cross-section and  $u_{\mathrm{CMB}} = 0.062\ \mathrm{
eV/cm}^3$  is the present energy density of the CMB. It is convenient
to write the solution in the inverted form: $E_0 = (-\kappa\ell +
1/E(\ell))^{-1}$ for substitution into $dN/dE$.  Numerically, we find
that the losses outside the diffusion zone make a small correction,
and that the distinction between $E_0$ and $E(\ell)$ is not important
here.

Each subhalo is characterized by a density profile that has been
fit to the Einasto form 
\be
\rho_i = \rho_{s,i}\exp\left[-{2\over\alpha}
	\left( \left(r\over r_{s,i}\right)^\alpha -1\right)\right]
\ee
with $\alpha = 0.17$ \cite{kuhlen}.  The scale radius is found to
be proportional to the radius $r_{v_{\rm max}}$ at which the velocity
dispersion is at a maximum, through the relation $r_s \cong 
r_{v_{\rm max}}/2.212$, while the prefactor scales with the maximum
velocity $v_{\rm max}$ as $\rho_s \cong v_{\rm max}^2 / (0.897\cdot
4\pi r_s^2 G)$.

\begin{figure}
\centerline{\includegraphics[width=0.4\textwidth]{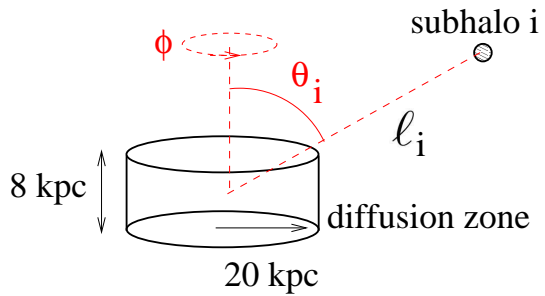}}
\caption{Geometry of a subhalo shining leptons on the boundary of diffusion zone of
the galaxy.}
\label{zone}
\end{figure}

To incorporate the contribution (\ref{flux}) to the lepton flux from
the subhalos in GALPROP, we add delta function source terms to
$q(r,z,p)$ for the cylindrical surface bounding the diffusion zone,
as illustrated in fig.\ \ref{zone}:
\bqa
	q_{\rm disk} &=& 2\delta(z\pm h/2)\sum_i {d\Phi_i\over
	dE}\cos\theta_i \nonumber\\
	q_{\rm band} &=& 2\delta(r-R)\sum_i {d\Phi_i\over
	dE}\sin\theta_i 
\eqa
where $h$ and $R$ are respectively the height and radius  of the 
cylinder.  The factor of 2 corrects for the fact that sources in
GALPROP have no directionality, whereas the flux impinging on the
surface is inward.  The sum is over the 20,048 resolved subhalos
in the \vl simulation.  In addition, the sources were averaged over
the azimuthal angle $\phi$ because GALPROP assumes cylindrical
symmetry in its 2D mode (and the 3D mode runs too slowly for our
purposes).  Finally, the distance $\ell_i$ must be corrected for
subhalos that are close to the diffusion zone; rather than the
distance to the center of the galaxy, it should be the distance to the
cylindrical boundary.  On average, this is a reduction by 17 kpc
compared to the distance to the galactic center.

Although most subhalos were located outside of the diffusion zone,
there are 143 lying inside, whose contribution required special
treatment.  Assuming approximate isotropy, we took their entire flux
of $e^++e^-$ to be pumped  into the diffusion zone from the boundary
rather than from their individual positions. Treating them in this
manner allowed us to group their contribution with that of the other
subhalos and thus consider a single average boost factor for all
subhalos.  This approximation would break down if one subhalo
happened to be very close to our position in the galaxy, but treating
such a case would anyway require going beyond the standard
cylindrical symmetry (2D) mode of  GALPROP and using the much slower
3D mode.  We believe this treatment is conservative in the sense that
it should only underestimate the contributions of the nearby
subhalos.

\section{Numerical Results}
\label{results}

We compared the observed flux of positrons and electrons generated by
dark matter annihilation within the main halo (MH) to a scenario in
which both main halo and subhalo (MH+SH) DM annihilation occurs. We
restricted our analysis to the DM DM $\rightarrow$ 4e channel. This
channel is simpler to analyze, and is somewhat less constrained 
by inverse Compton gamma ray constraints than the other 
4-lepton final-state models, or those with only two leptons
\cite{Meade}.

For the MH only scenario, the annihilation cross-section was
augmented by a constant boost factor $\bmh$, representing the effect
of Sommerfeld  or some other kind of enhancement
\cite{Hisano:2004ds,Cirelli:2007xd,ArkaniHamed:2008qn,Pospelov:2008jd}. 
This was varied in order to find a best fit to each data set. A
similar approach was used in the case of MH+SH, where we varied the
MH and SH boost factors independently.  This is justified by the
expectation that Sommerfeld enhancement should be significantly
larger in the subhalos due to their lower velocity dispersions
\cite{ArkaniHamed:2008qn}.  The subhalo boost factor  $\bsh$ might
also have further contributions besides Sommerfeld enhancement, such
as the presence of unresolved subhalos that we do not take into
account \cite{kuhlen}, as well as substructure within the subhalos
themselves \cite{Bovy}.

We minimized the chi squared coefficient
\begin{equation}
 \chi^2 = \sum_i \frac{\left(\xi_{i,\,\rm exp}-\xi_{i,\,\rm model} \right)
^2}{\sigma_{i,\,\rm exp}^2},
\label{chi2}
\end{equation}
where the sum runs over the measured or predicted values of $\xi$,  which stands
for either $E^3 d(\psi_{e^+} + \psi_{e^-})/dE$ in the case of the 25
Fermi data points, or $\psi_{e^+}/(\psi_{e^+} + \psi_{e^-})$ in the
case of the PAMELA data,
and $\psi_{e^\pm}$ is the flux of electrons or positrons. When fitting to PAMELA we excluded the first
8 of 16 data points, following the usual assumption that the dip relative to the background
is accounted for by solar modulation effects \cite{Watson}.

\subsection{Freely-varying background}
\label{sec:unconstrained}

Our best fits to the PAMELA and Fermi data were obtained by letting
the astrophysical background electrons and positrons be rescaled by
overall normalization factors, which was also the approach taken in
references \cite{Meade,Papucci:2009gd,Cirelli:2009dv}.    Adding subhalo contributions significantly
improved the fits to both the Fermi and the PAMELA data. While the
best overall fit with only MH electrons was for $\mchi = 850$ GeV
($\chi^2_{\rm total} = 34.3$), the MH+SH scenario gave a best fit at
$\mchi = 1.2$ TeV ($\chi^2_{\rm total} = 16.5$). In this case 31\% of
the DM electron + positron flux at the sun's location originated from
subhalos. 

\begin{figure}
\centerline{\includegraphics[width=0.5\textwidth]{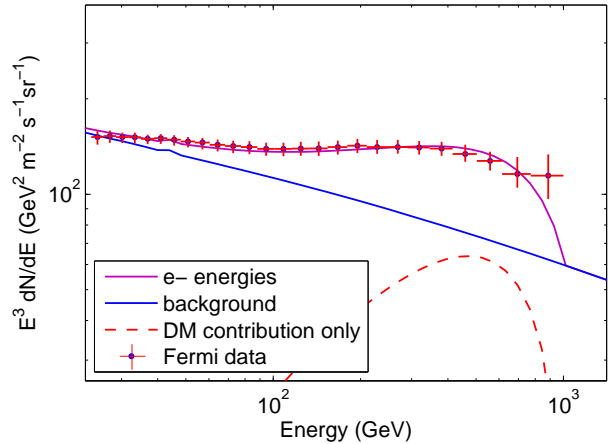}}
\caption{Fermi data and prediction for $e^++e^-$ flux 
of the best main-halo-only fit to
Fermi and PAMELA data, with an unconstrained background.}
\label{fermifitunMH} \end{figure}

\begin{figure}
\centerline{\includegraphics[width=0.5\textwidth]{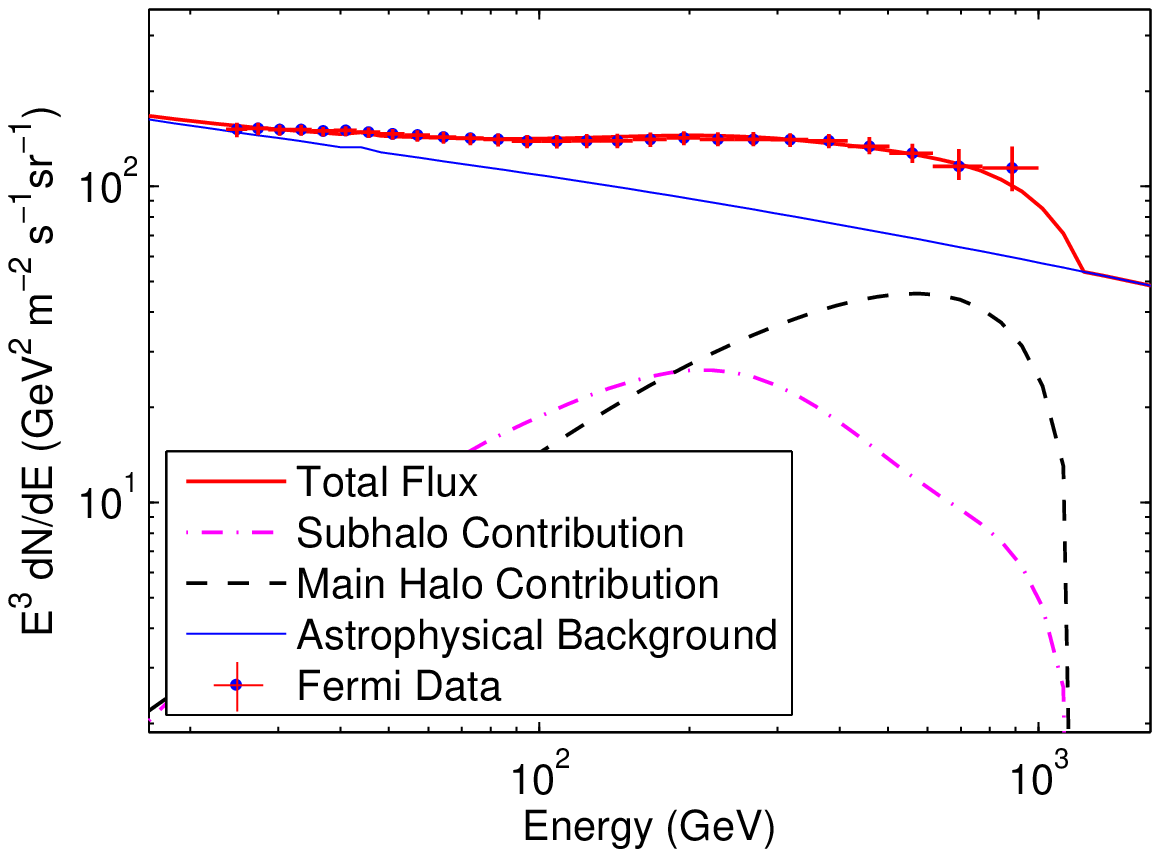}}
\caption{Same as fig.\ \ref{fermifitunMH}, but now including subhalo
contributions to the lepton fluxes.} \label{fermifitunSH} \end{figure}

\begin{figure}
\centerline{\includegraphics[width=0.5\textwidth]{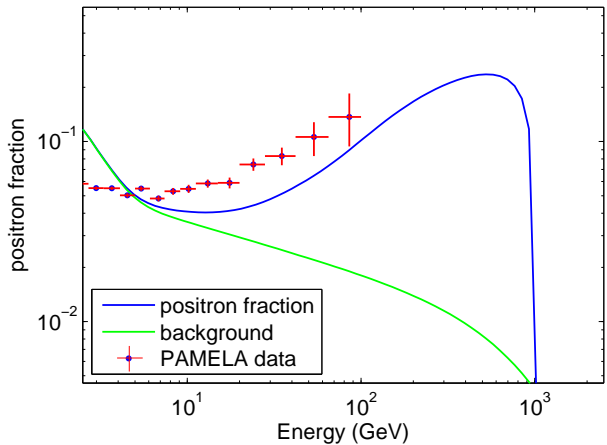}}
\caption{PAMELA data and predicted positron fraction of the 
best main-halo-only fit to Fermi and PAMELA data, 
with an unconstrained background.}
\label{pamfitunMH}
\end{figure}

\begin{figure}
\centerline{\includegraphics[width=0.5\textwidth]{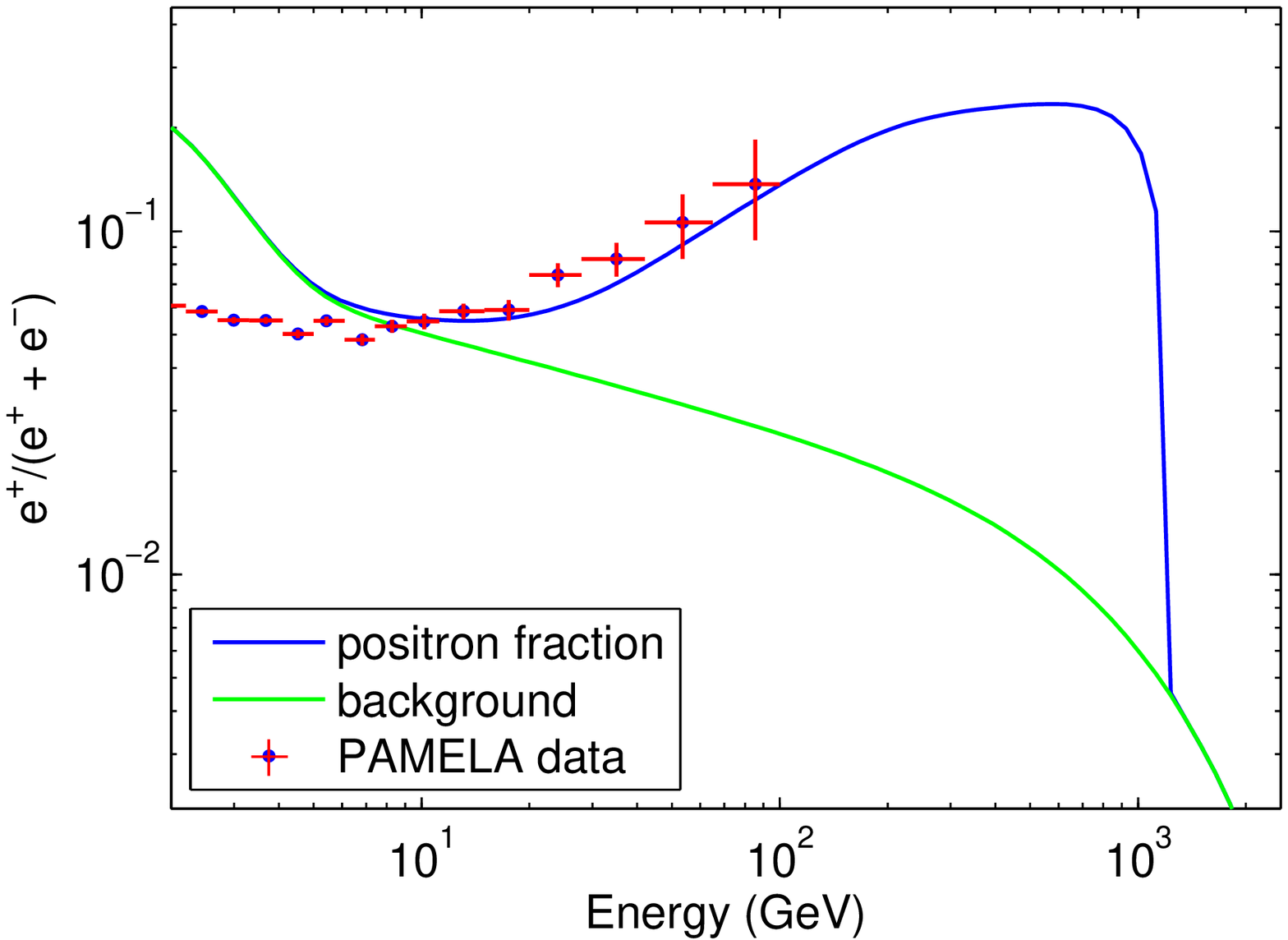}}
\caption{Same as fig.\ \ref{pamfitunMH}, but now including
subhalo contributions to the lepton fluxes.} \label{pamfitunSH} \end{figure}

\begin{figure}
\centerline{\includegraphics[width=0.5\textwidth]{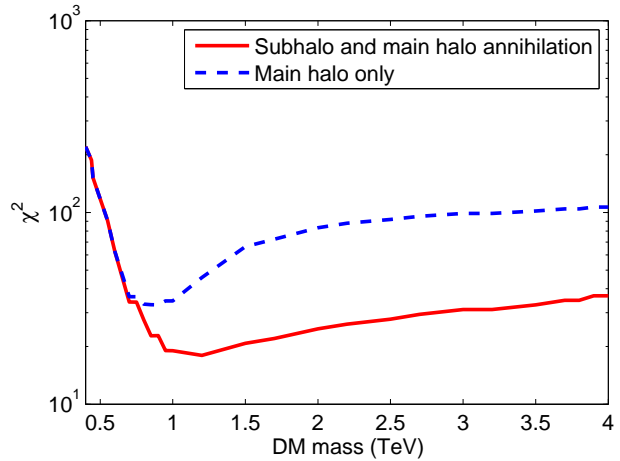}}
\caption{Combined $\chi^2$ for the Fermi and PAMELA data
as a function of the dark matter mass, for the unconstrained
background. Dashed (blue) line: main halo DM annihilation
only. Solid (red) line: subhalo and main halo
contributions combined.} \label{c2both} \end{figure}

A summary of these results is presented in the top portion
of Table \ref{resultTable}.   The predictions for MH and MH+SH
scenarios are shown for the total $e^++e^-$ flux in figures 
\ref{fermifitunMH} and \ref{fermifitunSH} and for the positron
fraction in figures \ref{pamfitunMH} and \ref{pamfitunSH}.
The value of $\chi^2$ versus $\mchi$ is shown in fig.\ \ref{c2both},
marginalizing over the background normalizations.
For the minimum $\chi^2$ point of the MH+SH model, the background
electrons had to be reduced to 97\% of their predicted values, while
background positrons were rescaled to 137\% of the GALPROP output.

The optimal boost factors of 90 for the main halo and 3800 for  the
subhalos are quite reasonable from the point of view of DM models that
give Sommerfeld-enhanced cross sections  \cite{ArkaniHamed:2008qn}. 
We leave for future work the issue of detailed particle physics model
building to match these and other features of the best-fitting models.

\begin{table}
 \begin{tabular}{|l|l|l|l|l|l|l|}
\hline
 \multicolumn{7}{|c|}{Freely-varying background} \\
 \hline
  & $\mchi$ & $\chi^2_{\rm Fermi}$ & 
$\chi^2_{\rm\scriptscriptstyle PAMELA}$ & $\chi^2_{\rm total}
$&$\bmh$& $\bsh$ \\ \hline
MH & 0.85 TeV & 15.5 & 18.7 & 34.3 & 90.3 & \ \ $-$\\
MH+SH & 1.2 TeV & 2.3 & 14.2 & 16.5 &92.8 & 3774 \\
\hline 
\multicolumn{7}{|c|}{Fixed GALPROP background} \\
\hline
MH & 1.0 TeV & 8.2 & 144 & 152.2 & 110 & \ \ $-$\\
MH+SH & 2.2 TeV & 2.1 & 175 & 177.1 &146 & 1946 \\
\hline
\end{tabular}
\caption{Best fit scenarios. Top: when the background positron and
electron spectra were allowed to vary by an overall factor; this
corresponds to the best overall fit to the data. Bottom: using
background that was fixed at GALPROP's normalization. In this case we
used the best fit to Fermi, since the best overall fit gave values of
electron + positron flux that were ruled out by the Fermi data. MH:
main halo DM annihilation only. MH+SH: subhalo annihilation included.
$m_{DM}$ is the DM mass that gives the best fit and $\chi^2_i$ are
the chi squared fits to the respective experiments as described in
eq. (\ref{chi2}). $\bmh$ and $\bsh$ are the boost factors
necessary for MH and SH annihilation cross sections, respectively. Note
that the addition of a subhalo contribution greatly improves the best
fit for both Fermi and PAMELA. The required DM mass is larger because
of the energy loss suffered by electrons propagating to us
from the galactic edge.} \label{resultTable} \end{table}

\subsection{Constrained background}
\label{sec:constrained}

We performed a second analysis by taking the electron and positron
backgrounds to be those predicted by GALPROP. In this case, although
there is no good simultaneous fit to the combined PAMELA and Fermi
data, we nevertheless find that SH contributions improve the fit. In
rough agreement with ref.\ \cite{Cholis:2008qq}, we find that the
PAMELA data require a boost factor several times higher than that
needed to fit the Fermi data. 

\begin{figure}[t]
\centerline{\includegraphics[width=0.5\textwidth]{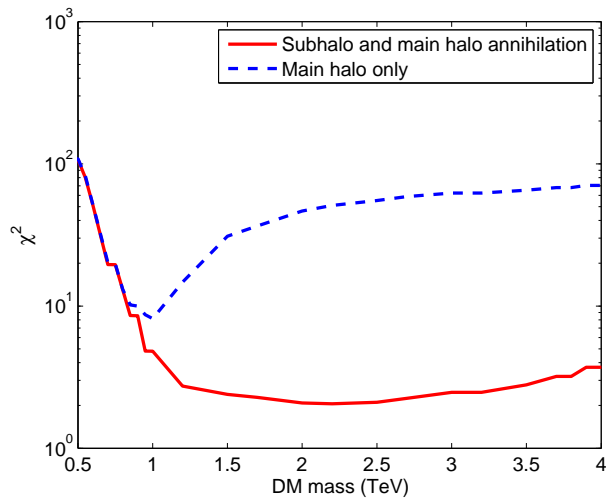}}
\caption{$\chi^2$ versus $\mchi$ for the 
Fermi $e^++e^-$ data
using the GALPROP constrained background.
Dashed (blue) line: main halo only.
Solid (red) line: subhalos plus main halo.} \label{c2f} \end{figure}

\begin{figure}[h]
\centerline{\includegraphics[width=0.5\textwidth]{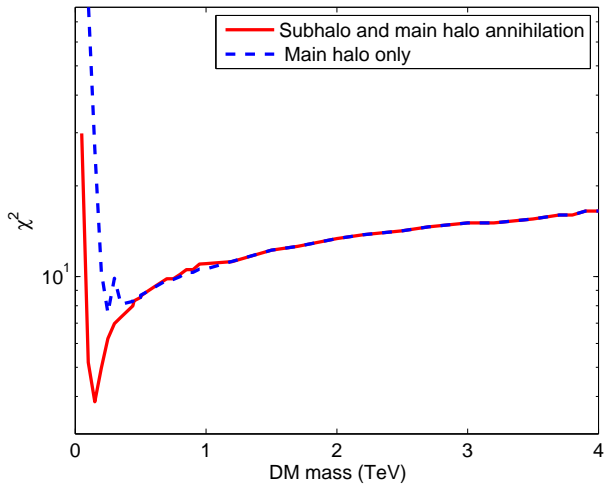}}
\caption{$\chi^2$ versus $\mchi$ for the 
Pamela positron fraction data,
using the GALPROP constrained background.
Dashed (blue) line: main halo only.
Solid (red) line: subhalos plus main halo.}
\label{c2p} \end{figure}

\begin{figure}[t]
\centering
\subfigure[]{
\includegraphics[scale=.623]{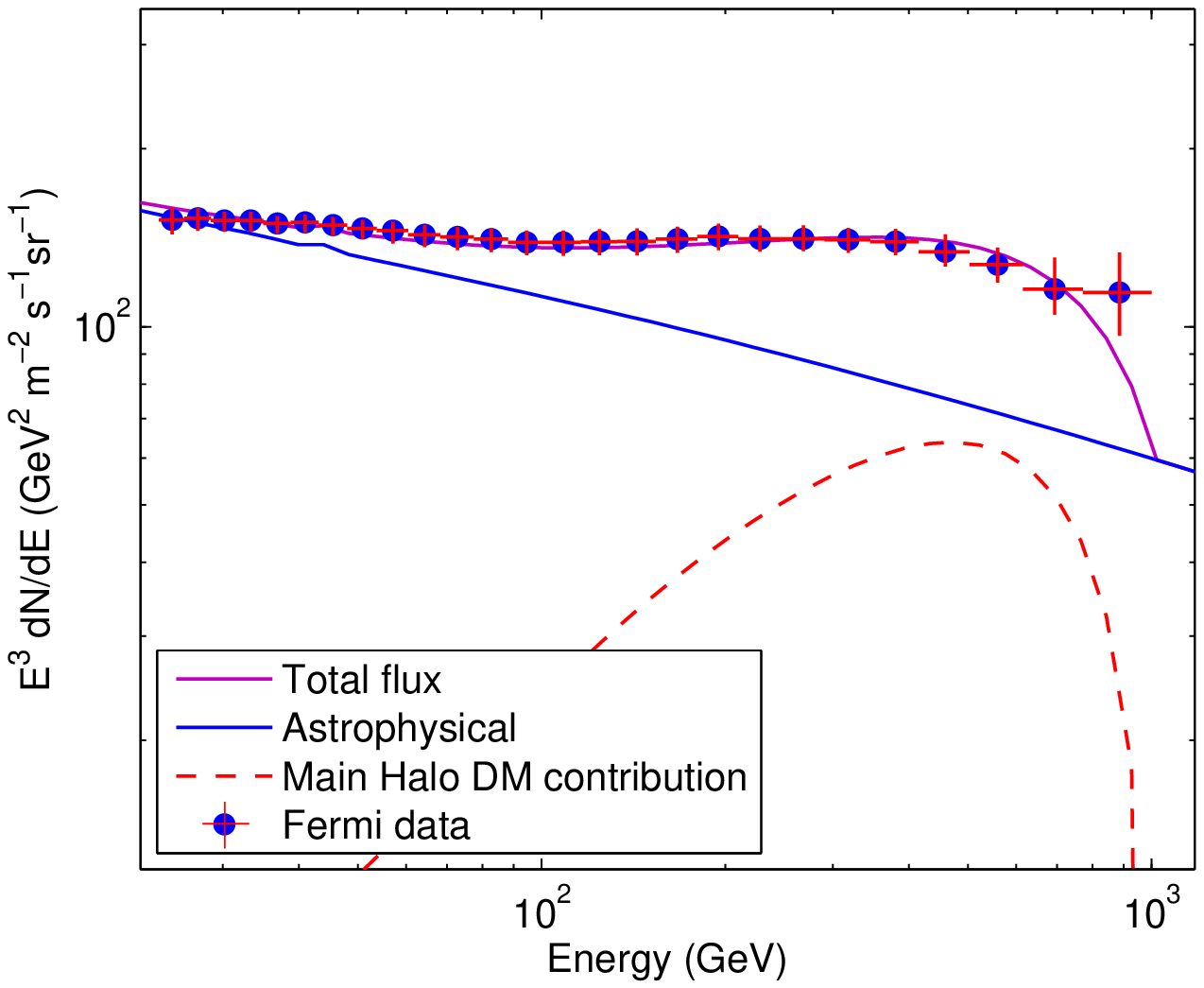}
\label{fig:subfig1}
}
\subfigure[]{
\includegraphics[scale=.58]{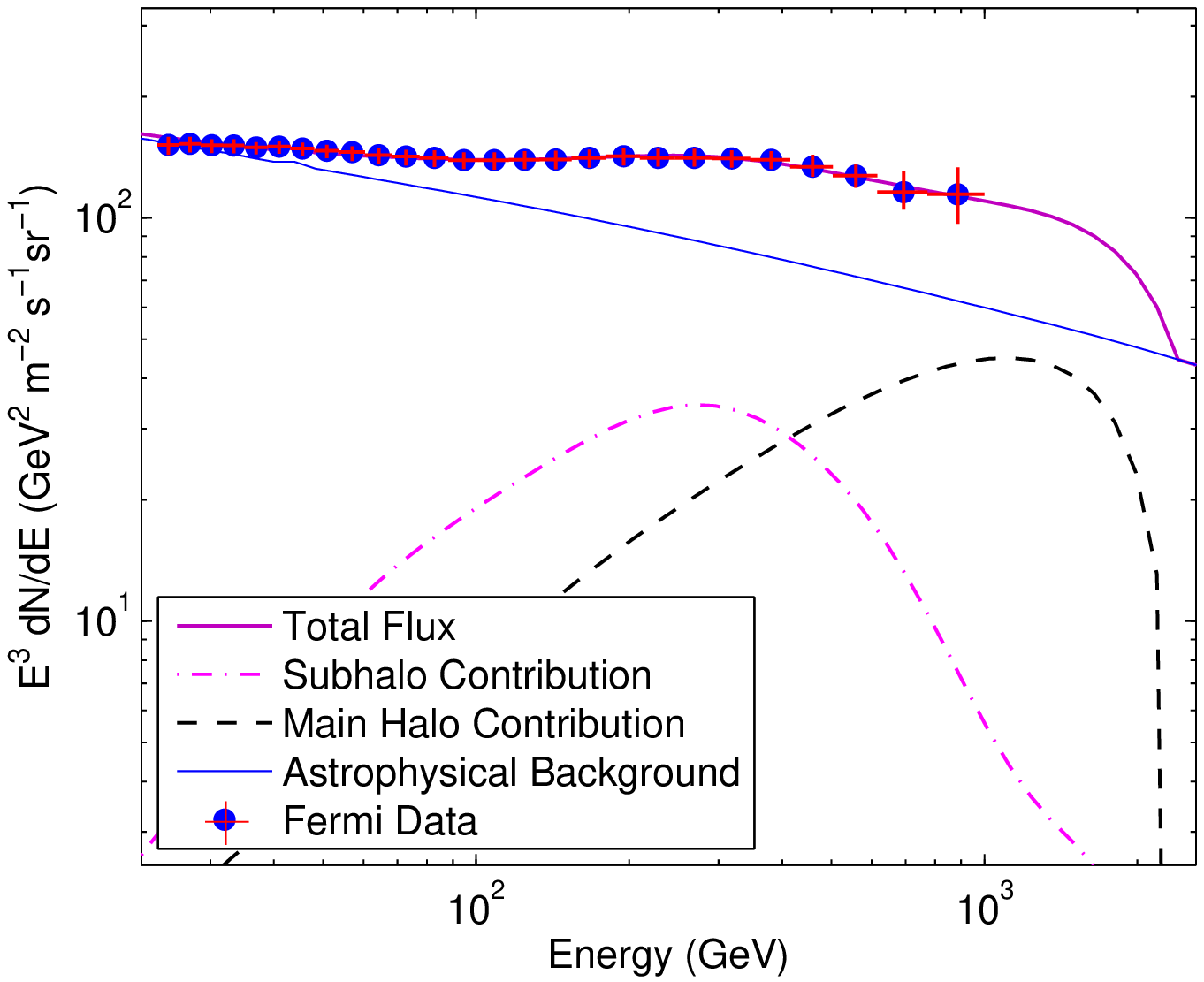}
\label{fig:subfig2}
}
\caption{Best fits to Fermi data. a) Main halo only,
 $\mchi = 1$ TeV, with a boost factor $\bmh =
110$. b) Subhalos plus main halo, $\mchi = 2.2$ TeV
and $\bmh = 146$, $\bsh = 4825$.  } \label{bestfitsfig}
\end{figure}

\begin{figure}[t]
\centerline{\includegraphics[width=0.5\textwidth]{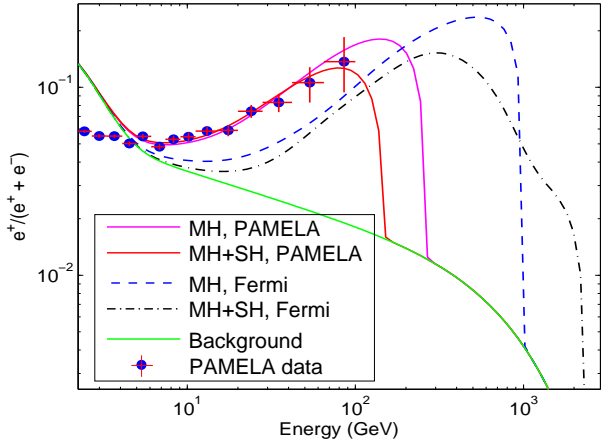}}

\caption{Positron fraction for four of the best fit scenarios with
constrained backgrounds. Top
(solid) lines correspond to fits to PAMELA data only. Uppermost
(magenta): MH
only; $m_{DM}$ = 250 GeV, $S_{MH}$ = 225. Lower (red): MH+SH; $m_{DM}$ = 150
GeV, $S_{MH}$ = 9.3, $S_{SH}$ = 509. Bottom (dashed) lines correspond
to the best fits of these scenarios to the Fermi data. Upper dashed
(blue):
MH only; $m_{DM}$ = 1 TeV, $S_{MH}$ = 110. Lower dot-dashed (black): MH+SH;
$m_{DM}$ = 2.2 TeV, $S_{MH}$ = 146, $S_{SH}$ = 4825. Although the
former set provide a better $\chi^2$, they predict a total 
$e^++e^-$ flux
that conflicts with the Fermi data by at least 3$\sigma$  (see figure
\ref{badfit}).} \label{pamfig} \end{figure}
\begin{figure}[!]
\centerline{\includegraphics[width=0.5\textwidth]{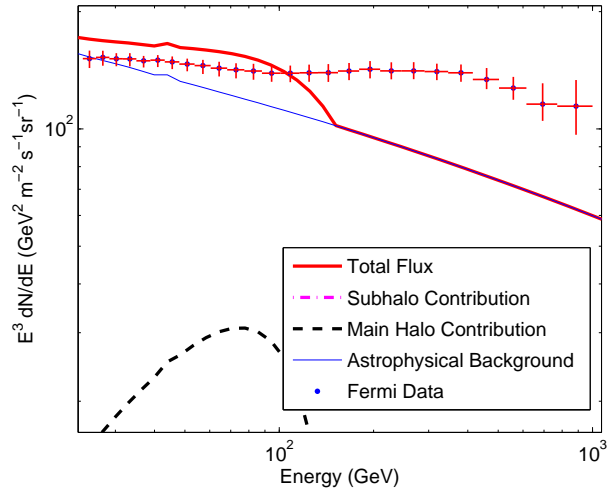}}
\caption{e$^+$ + e$^-$ curve for best fit to PAMELA data only. $m_{DM} = 150$ GeV, $S_{MH} = 9.3$, $S_{SH} = 509$. Although the high energy tail could be compensated by other sources (e.g. pulsars), the fact that the model exceeds the data points at low energy leads us to disfavor this model. Note that the subhalo contribution is too small to be seen in this figure.}
\label{badfit}
\end{figure}

The plots of $\chi^2$ versus $\mchi$, for both the MH-only and MH+SH
models, are shown respectively for the Fermi and PAMELA data in
figures \ref{c2f} and \ref{c2p}. It is striking that the best-fit DM
mass becomes significantly larger and less constrained in the fit to
the Fermi data including subhalos, fig. \ref{c2p}. The increase in
the required DM mass is due to the energy lost by electrons and
positrons during propagation from the edge of the diffusion zone to
our position. The best fit to the Fermi data has $\mchi  = 2.2$
TeV ($\chi^2 = 2.05$) with SH+MH, compared with $\mchi = 1$ TeV
($\chi^2 = 8.15$) in the MH only case. The required boost factors for
these fits are $\bmh = 146$ and $\bsh = 1946$ for SH+MH, in
contrast with $\bmh = 110$ for MH only.  The best fit cases for the
$e^++e^-$ spectrum are
shown in Figure \ref{bestfitsfig} and the results are summarized in the
bottom part of Table \ref{resultTable}. The corresponding positron
fraction in each of these scenarios is shown in Figure
\ref{pamfig}.

The fit to PAMELA is also improved by the addition of SH positrons,
but only at low DM mass, $\mchi< 500$ GeV. However the best fit
parameters for the PAMELA data by themselves lead to a prediction of
the $e^++e^-$ $E^3 dN/dE$  spectrum that exceeds the Fermi data  by
more than 3$\sigma$, resulting in a $\chi^2$ = 460 fit to Fermi.  The
badness of this fit is evident in fig.\ \ref{badfit}.  

\subsection{Relative contributions of subhalos}
\begin{figure}
\centerline{\includegraphics[width=0.5\textwidth]{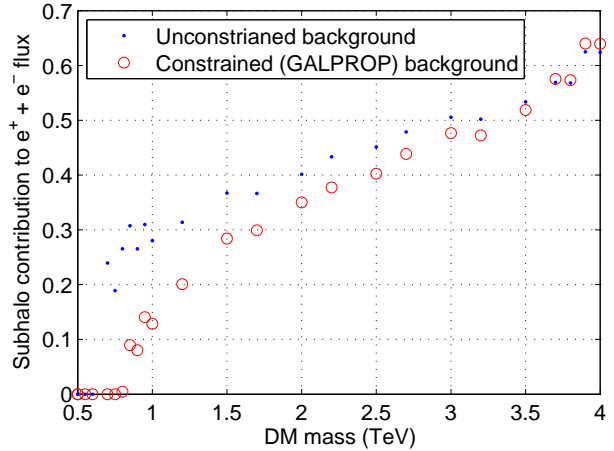}}
\caption{Proportion of the total flux of $e^++e^-$ originating
from subhalos as opposed to the main halo, 
as observed 8.5 kpc  from the
galactic center (the position of the solar system)
in order to obtain a best fit to the Fermi data. Each
point represents an individual simulation.} \label{fraction}
\end{figure}

It is interesting to quantify how much of the signal can be
contributed by the subhalos relative to that coming from the
main halo.  We show the fraction of $e^++e^-$ pairs due to the
subhalos, as a function of the DM mass, in fig.\ \ref{fraction}.
For the best-fit values of the mass, this fraction is around 30\%,
but for larger values of $\mchi$ (yet still giving reasonable fits)
it rises to 60\% or more.  This may be helpful for weakening the
constraints on the model from production of gamma rays by inverse
Compton scattering \cite{Cirelli:2009dv,Papucci:2009gd}.  We hope to 
investigate this issue in the near future.

\begin{figure}[b]
\centerline{\includegraphics[width=0.5\textwidth]{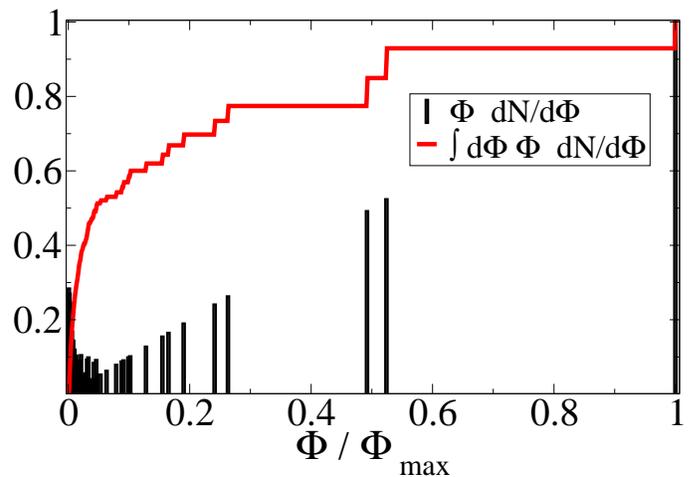}}
\caption{Vertical bars (black): $\Phi\, dN/\d\Phi$, the distribution of subhalos
contributing a given flux at the edge of the difusion zone, weighted
by the flux.  Continuous curve (red): integral of $\Phi\, dN/\d\Phi$.} \label{hist}
\end{figure}

Another relevant issue is the hierarchy of contributions of subhalos
relative to each other.  One would like to know whether it was really
necessary to add the contributions of all 20,000 subhalos, or if
perhaps only the few closest ones dominate.  Fig.\ \ref{hist} shows
the distribution of subhalos contributing a given flux $\Phi$
(normalized to the contribution of the subhalo that gives the largest
value $\Phi_{\rm max}$), weighted by the flux, and also the integral
of this quantity.  From the integral, we see that 50\% of the total
signal comes from subhalos whose individual intensities are less than 
5\% of the strongest one.  Thus to get a quantitatively accurate
estimate, it is necessary to include the very numerous subhalos whose
intensity is low.  This also suggests that our computation is an
underestimate, since we do not count the subhalos that are not
resolved by the \vl simulation.

\section{Conclusions}
\label{conclusion}

We have shown that the inclusion of electrons and positrons  from the
galactic subhalos can  significantly alter the predictions from
annihilating dark matter models. Using the \vl simulated data of the
subhalo distribution around a Milky Way-like galaxy, we found that
the contributions from substructure can give improved fits to the
PAMELA and Fermi excess lepton data, and increase the value of the 
expected mass of the dark matter particle. A strong Sommerfeld boost
coming from the low velocity dispersions of the subhalos, as well as
the uncounted  contributions of subhalos unresolved by the \vl
simulations, are possible sources for the boost factor necessary to
obtain best our best fits to the data.  According to these fits a
third or more  of the electron cosmic rays from DM annihilation could
come from subhalos outside of the visible Milky Way.

The next step for future work will be to see whether the reduction of
the flux from the main halo can weaken Fermi constraints on
annihilating DM models due to the inverse Compton gamma rays produced
by the high-energy leptons \cite{Cirelli:2009dv,Papucci:2009gd}.
These constraints are sufficiently strong to rule out the DM
interpretation of the Fermi lepton excess, under the usual assumption
that all the $e^+ e^-$ pairs are produced in the main halo.  The
constraints are strongest from data near the galactic center.  By
shifting the production away from the center to the subhalos, the
constraints should be weakened, but whether the effect is large
enough to reinstate the DM interpretation of the Fermi lepton
observations is a quantitative question.  In addition, one should
satisfy other  protohalo constraints \cite{Francis}, extragalactic
gamma background \cite{Huetsi:2009ex} and last scattering surface CMB
constraints \cite{Slatyer:2009yq}.

If it is possible for the scenario to pass these tests, it will be
interesting to check whether specific particle physics models are
able to give the average boost factors  that we have treated as free
parameters in this preliminary study.  

As we were completing this work, ref.\ \cite{Kamionkowski:2010mi}
appeared, which presents an analytical method for taking into account
the effect of substructure on dark matter annihilation.

\section*{Acknowledgments}  We thank Ilias Cholis, Ran Lu, Scott
 Watson and Neal Weiner for helpful discussions and correspondence concerning
 GALPROP. Our research is supported by NSERC (Canada) and FQRNT
 (Qu\'ebec).

\bibliographystyle{h-physrev}
\bibliography{galprop.bib}

\end{document}